\documentclass{aastex701}

\received{August 29, 2026}
\revised{September 27, 2026}
\accepted{October 11, 2026}
\submitjournal{ApJL}

\begin{document}

\title{The origin of isolated millisecond pulsars in non-core-collapsed globular clusters}

\author[orcid=0000-0001-5489-4925,sname='de Menezes']{Raniere de Menezes}
\affiliation{Centro Brasileiro de Pesquisas F\'isicas, 22290-180, Rio de Janeiro, RJ, Brazil}
\email[show]{raniere@cbpf.br}

\begin{abstract}

A significant fraction of millisecond pulsars (MSPs) in globular clusters (GCs) are observed as isolated objects, despite the widely accepted scenario in which MSPs are formed through recycling in compact binary systems. The origin of these isolated objects therefore remains an open problem. In this work, we propose a combination of the Heggie--Hills law for hard/soft binaries with companion ablation as a tracer of the fraction of isolated MSPs, $\mathcal{F}_i$, observed in non-core-collapsed GCs. We find $\mathcal{F}_i = (42.49\pm1.62)a_H^{-0.5}\,\%$, where $a_H$ is the Heggie--Hills ionization radius, and that this ionization-driven model is favored over a null hypothesis in which $\mathcal{F}_i$ is independent of $a_H$, with a log Bayes factor of $\ln(BF) > 2.5$, providing significant evidence for a physical dependence of $\mathcal{F}_i$ on $a_H$. Our framework naturally explains the observed overabundance of isolated MSPs in NGC 5139 (aka $\omega$ Centauri) and proposes binary ionization as one of the main mechanisms responsible for the production of isolated MSPs in non-core-collapsed GCs.

\end{abstract}

\keywords{\uat{Globular star clusters}{656} --- \uat{Millisecond pulsars}{1062} --- \uat{Compact binary stars}{283}}


\section{Introduction}
\label{sec:intro} 

Young pulsars, such as the Crab and Vela, lose rotational energy at a rate $\dot{E}_{\rm rot} = -4\pi^2 I \dot{P} / P^3$ \citep{lorimer2005handbook}, where $I$ is the moment of inertia, $P$ the rotation period, and $\dot{P}$ its time derivative. Over the course of $10^7$--$10^8$ years, these pulsars fade away, becoming ordinary neutron stars, as the voltage in their magnetospheres \citep[i.e., $\Delta V \propto B/P^2$, where $B$ is the surface magnetic field;][]{goldreich1969pulsar} falls below the pair production threshold. However, if these neutron stars reside in dense stellar environments, such as the core of a globular cluster (GC), where number densities can easily exceed $10^3$ stars pc$^{-3}$ \citep{sollima2017global}, gravitational encounters can capture them into compact binary systems. In particular, low-mass X-ray binaries (LMXBs) allow neutron stars to accrete matter and be spun up, eventually evolving into millisecond pulsars (MSPs) bound in binary systems.

In an LMXB, the accretion rate onto the neutron star is limited by the Eddington threshold to $\lesssim 10^{-8}$ M$_\odot$ yr$^{-1}$. Accretion of as little as $\sim 0.1$ M$_\odot$ of material is already sufficient to accelerate the neutron star to millisecond periods \citep{verbunt1987formation}. The MSP phase begins when mass transfer ceases or declines sufficiently for the magnetosphere to re-expand \citep{archibald2009radio}, which occurs when the magnetospheric radius equals the light cylinder radius,
\[
r_m = \left( \frac{\mu^4}{2GM\dot{M}^2} \right)^{1/7} = r_{\rm LC} = \frac{cP}{2\pi},
\]
where $\mu$, $P$, $M$, and $\dot{M}$ are the neutron star's magnetic dipole moment, rotation period, mass, and accretion rate, respectively; $G$ is the gravitational constant, and $c$ is the speed of light \citep[see, e.g., Chapter 6 in][]{frank2002accretion}. Given that typical stellar masses in GCs are $m_{\rm typ} \sim 0.3$--$0.4$ M$_\odot$ \citep{paresce2000globular}, the companion star at the birth of an MSP is expected to have a mass roughly in the range $0.2$--$0.3$ M$_\odot$.

This formation channel for MSPs is supported by several observations showing a linear correlation between the stellar encounter rate $\Gamma$ and either the number of X-ray sources $N_X$ or the gamma-ray luminosity in the cores of GCs \citep{pooley2003dynamical,bahramian2013stellar,demenezes2019milky,evans2023gamma}. However, \citet{deMenezes2023dynamical} demonstrate that while the $N_X \propto \Gamma$ correlation holds for most clusters, it breaks down for $\Gamma \lesssim 100$ (on a scale where NGC 104, aka 47 Tucanae, has $\Gamma \equiv 1000$). In this regime, clusters such as NGC 5139 (aka $\omega$ Centauri, with $\Gamma \approx 90$) host more X-ray sources than predicted by the linear correlation. Furthermore, the specific abundances of LMXBs and MSPs in GCs are roughly two orders of magnitude higher than in the Galactic field \citep{katz1975two,smith2023third,Kremer2026_compact_objects_in_GCs}, suggesting that dynamical processes unique to these dense environments provide efficient pathways for forming MSPs. Modern N-body simulations \citep{ye2019millisecond} further confirm that dynamical interactions play a critical role in MSP formation in GCs. Taken together, these results support a scenario in which dynamical formation is one of the most important MSP formation channels \citep[see, e.g.,][]{verbunt1987formation,Sigurdsson1995Dynamics,ivanova2008Formation,ye2022_compact}.

The major remaining challenge in this scenario is understanding the large population of isolated MSPs observed in GCs \citep[e.g.,][]{dai2020discovery,chen2023meerkat,ye2024Single_Millisecond}. \citet{verbunt1987formation} proposed that a direct collision between the companion star and a nearby star could disrupt the binary, leaving the MSP with a temporary massive accretion disk. In subsequent work \citep{verbunt2014disruption}, the same author suggested that, after the MSP forms, secondary encounters between the compact binary and nearby stars can ionize the system at a rate
\begin{equation}
    \gamma \approx \frac{2\pi GM_{tot}n a}{\sigma},
    \label{eq:enc_rate_per_binary}
\end{equation}
where $n$ is the local stellar number density, $M_{tot}$ is the sum of the masses of the three stars involved in the interaction, $a$ is the binary's semi-major axis, and $\sigma$ is the velocity dispersion of nearby stars. This ionization scenario has also been investigated in the context of core-collapsed clusters: \citet{oh2023influences} found that estimated disruption rates and isolated-pulsar fractions are predominantly higher in core-collapsed GCs, whereas \citet{ye2024Single_Millisecond} used N-body simulations to show that binary disruption is ineffective as an explanation for the excess of single pulsars in the core-collapsed cluster NGC 6752.

Additional channels may operate specifically for isolated MSPs, especially in core-collapsed GCs, including the merger of heavy white dwarf binary systems, which can collapse into a neutron star \citep{ye2024white_dwarf_collision}, and the tidal disruption of main-sequence stars in close encounters with neutron stars \citep{camilo2005psr_in_gc,kremer2022formation_of_low-mass}. In the latter case, the neutron star can accrete part of the bound debris and thereby reach millisecond rotation periods.

In this work, we argue that the Heggie-Hills law \citep{heggie1975binary,hills1975encounters} for hard/soft binaries, combined with companion mass loss driven by intense MSP winds and radiation pressure \citep[e.g., the black widow systems;][]{gentile2014xray_black_widow}, gives a robust prediction of the fraction of isolated MSPs, $\mathcal{F}_i$, observed in non-core-collapsed GCs. This is primarily because the Heggie-Hills critical semi-major axis, $a_H$, can be determined by global properties of the GC (i.e., stellar velocity dispersion). This combined approach yields a predictive model for the observed fraction of isolated MSPs in non-core-collapsed GCs.

\section{The proposed scenario}
\label{sec:scenario}

\begin{figure*}
    \centering
    \includegraphics[width=\linewidth]{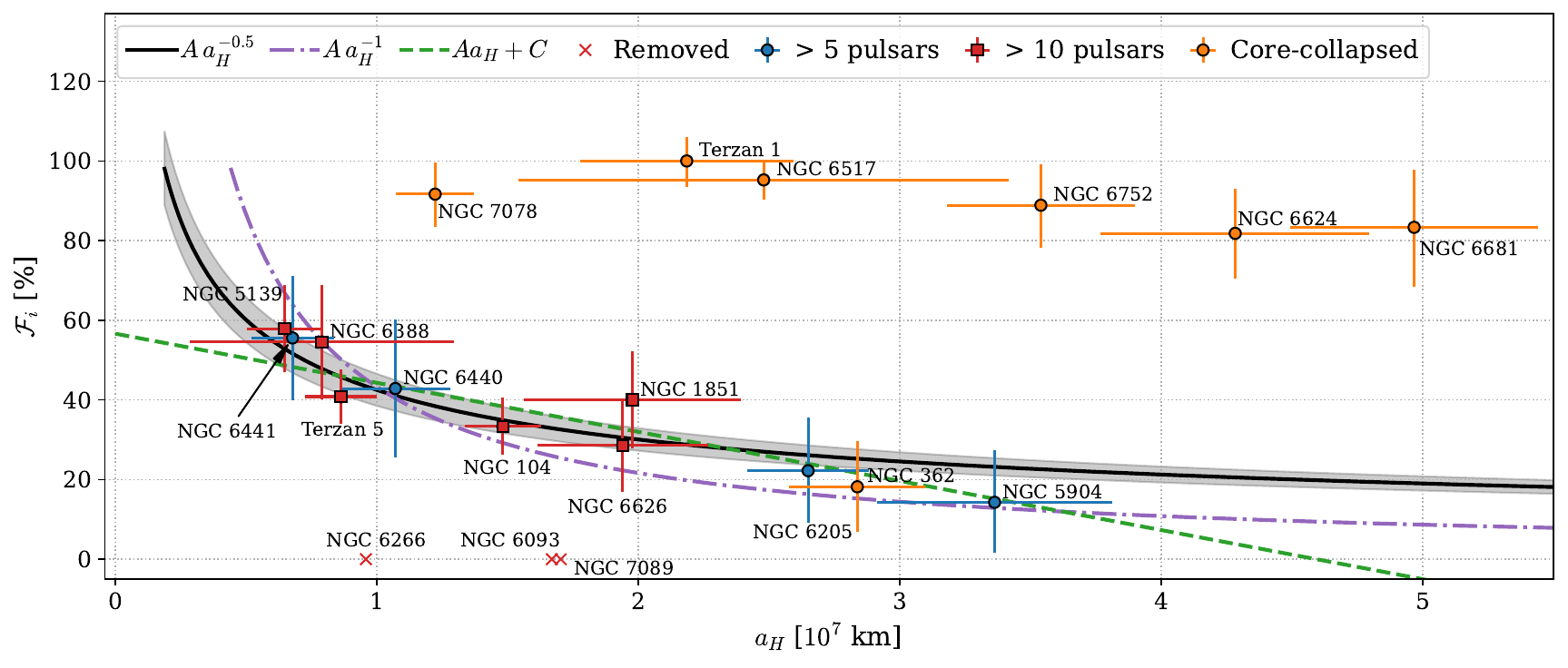}
    \caption{Fraction of isolated MSPs in GCs as a function of the threshold orbital separation $a_H$. Smaller values of $a_H$ shift the hard--soft boundary to smaller orbital separations, so that even relatively compact binaries fall in the soft, ionizable regime (i.e., $a \gtrsim a_H$) and are disrupted through encounters. Non-core-collapsed clusters with more than 10 detected MSPs are shown in red, while those with more than 5 are shown in blue. Lines of different colors show the three simplest models tested (see Table~\ref{tab:odr_fits}), each fitted to the non-core-collapsed clusters (red and blue points); the shaded grey region marks the $1\sigma$ confidence interval of our fiducial model, $A\,a_H^{-0.5}$ (black line).}
    \label{fig:Frac_x_a_H}
\end{figure*}

The Heggie-Hills law \citep{heggie1975binary,hills1975encounters} governs the dynamical evolution of binaries in dense stellar environments such as GC cores: hard binaries become harder, while soft binaries become softer. A binary is classified as hard when its binding energy greatly exceeds the typical kinetic energy of surrounding stars,
\[
|E_b| = \frac{G m^2}{2a} \gg \frac{m\sigma^2}{2},
\]
and soft when the inequality is reversed,
\[
|E_b| = \frac{G m^2}{2a} \ll \frac{m\sigma^2}{2},
\]
where $G$ is the gravitational constant, $m$ is the characteristic stellar mass (assuming that all stars have the same mass), $\sigma$ is the velocity dispersion, and $a$ is the orbital semi-major axis. Binaries with $|E_b| \sim m\sigma^2/2$ occupy a transitional regime where encounters can either harden or soften them, defining a characteristic separation scale at which ionization becomes possible. This scale can be estimated independently for each Milky Way GC.

However, assuming equal masses for all stars may be inadequate for our problem: a newly formed MSP can ablate its companion through relativistic particle winds, Poynting flux, and high-energy radiation \citep{fruchter1988millisecond,devito2020identifying,ginzburg2021black}. In extreme cases, known as black widow systems, the companion can be eroded to masses below $0.01$ M$_{\odot}$ \citep{romani2012psr,spiewak2018psr}. Accounting for these mass differences, we follow \citet{hut1983binary} to redefine the ionization threshold via the relation
\[
\frac{G M_{\rm NS} M_c}{2a_{H}} = \frac{\mu_3 v_{\infty}^2}{2},
\]
where $\mu_3$ is the reduced mass of the three-body system comprising the binary and an incident star, $v_{\infty}^2 = 3\sigma^2$ is the typical velocity of the incoming star at infinity relative to the binary's center of mass, and $a_H$ is the critical semi-major axis separating hard and soft regimes. Expressing $\mu_3$ in terms of the stellar masses and solving for $a_H$ yields
\[
\frac{G M_{\rm NS} M_c}{2a_H}
\simeq
\frac{1}{2}
\frac{m_\ast (M_{\rm NS} + M_c)}
{M_{\rm NS} + M_c + m_\ast}
\, 3\sigma^2,
\]

\begin{equation}
a_{H}
=
\frac{G M_{\rm NS} M_c}
{3\sigma^2
\displaystyle
\frac{m_\ast (M_{\rm NS} + M_c)}
{M_{\rm NS} + M_c + m_\ast}
},
\label{eq:a_H}
\end{equation}
where we adopt $M_{\rm NS} = 1.5$ M$_{\odot}$ for the neutron star mass, $M_c = 0.01$ M$_{\odot}$ for the companion mass, and $m_\ast \approx 0.4$ M$_{\odot}$ for the typical mass of surrounding stars. For this parameter set, the resulting $a_H$ values for Milky Way GCs lie in the range $5$--$50 \times 10^6$ km. This is already comparable to the typical separations in black widow systems, which from Kepler's third law,
\[
a \simeq \left( \frac{G M_{\rm NS}}{4\pi^2} \right)^{1/3} P_{\rm orb}^{2/3},
\]
fall in the range $0.4$--$2 \times 10^6$ km for orbital periods $P_{\rm orb} \sim 1$--$10$ hours. The similarity between these scales implies that compact binaries with low-mass companions in GCs are already susceptible to dynamical ionization. Furthermore, adopting even smaller companion masses, as observed in some systems \citep{spiewak2018psr}, or larger values of $m_\ast > 0.4$~M$_\odot$ (as expected in GC cores, where mass segregation via dynamical friction concentrates the heavier stars) can produce a strong overlap between these scales.

In Figure~\ref{fig:Frac_x_a_H} (see below for details on the data), we show evidence that the fraction of isolated MSPs (relative to the total number of MSPs) in Milky Way non-core-collapsed GCs increases as $a_H$ decreases. This result supports binary ionization as a formation channel for isolated MSPs in non-core-collapsed GCs, and highlights that this process depends directly on the global properties of the GC (since $a_H$ depends on $\sigma$) and the ablation of the MSP companion.

The data used in Figure~\ref{fig:Frac_x_a_H} (i.e, the total number of binary and isolated MSPs and the nuclear velocity dispersions of the GCs) were compiled from the online databases maintained by Freire and Baumgardt\footnote{\url{https://www3.mpifr-bonn.mpg.de/staff/pfreire/GCpsr.html}}\footnote{\url{https://people.smp.uq.edu.au/HolgerBaumgardt/globular/}}. These compilations synthesize results from numerous independent studies \citep[e.g.,][]{freire2017long,baumgardt2018catalogue,sollima2019eye,dai2020discovery,pan2021fast,baumgardt2023evidence,ridolfi2022trapum,chen2023meerkat,padmanabh2024discovery,yin2025illuminating}. We define the isolated MSP fraction as $\mathcal{F}_i = N_i/N$, where $N_i$ is the number of isolated MSPs and $N$ is the total number of MSPs in a given GC. We consider all pulsars with spin periods $P < 100$ ms to be MSPs, and exclude slower pulsars from our analysis. Adopting binomial statistics for this fraction, we compute uncertainties as Bayesian credible intervals using a Jeffreys prior for the binomial proportion $\mathcal{F}_i$, i.e., $\mathcal{P}(\mathcal{F}_i) \sim \mathrm{Beta}(1/2,1/2)$. For a given $N$ and $N_i$, this yields the posterior distribution
\[
p(\mathcal{F}_i \mid N, N_i) = \mathrm{Beta}\!\left(N_i+\frac{1}{2},\,N-N_i+\frac{1}{2}\right),
\]
which produces slightly asymmetric error bars. As a sensitivity check, we repeat the analysis adopting a uniform prior, $\mathcal{P}(\mathcal{F}_i) \sim \mathrm{Beta}(1,1)$, which yields the posterior $p(\mathcal{F}_i \mid N, N_i) = \mathrm{Beta}(N_i+1,\,N-N_i+1)$. Uncertainties in $a_H$ are propagated from the velocity dispersion via $\Delta a_H = 2a_H\Delta\sigma/\sigma$. The adopted $\sigma$ values are the one-dimensional central velocity dispersions estimated for each GC\footnote{\url{https://people.smp.uq.edu.au/HolgerBaumgardt/globular/parameter.html}}; the conversion to a three-dimensional (isotropic) velocity, via $v_\infty^2 = 3\sigma^2$, is applied directly in Eq.~\ref{eq:a_H}. The associated uncertainties are estimated as the average of the asymmetric upper and lower error bounds reported for the innermost bin of each cluster\footnote{\url{https://people.smp.uq.edu.au/HolgerBaumgardt/globular/veldis.html}}. These values are listed in Table~\ref{tab:velocity_dispersion}.

\begin{table}[h]
  \centering
  \caption{One-dimensional central velocity dispersions ($\sigma$) and central densities ($\rho_c$) adopted for each GC. Uncertainties in $\sigma$ are estimated as the average of the innermost upper and lower error bounds reported in the Baumgardt database. *For display purposes, the $\sigma$ value of NGC 6093 was shifted by 2\% in Figure~\ref{fig:Frac_x_a_H} so that the point does not overlap with NGC 7089.}
  \label{tab:velocity_dispersion}
  \vspace{0.5em}
  \begin{tabular}{lcc}
    \hline
    \textbf{Name} & \textbf{$\sigma$ [km/s]} & \textbf{$\log_{10}(\rho_c)$ [$M_\odot\,\mathrm{pc}^{-3}$]} \\
    \hline
    \multicolumn{3}{l}{\textit{Main sample}} \\
    \hline
    NGC 104   & $11.9 \pm 0.6$ & $4.72$ \\[0.3em]
    NGC 1851  & $10.3 \pm 1.1$ & $6.32$ \\[0.3em]
    NGC 5139  & $18.0 \pm 2.0$ & $3.23$ \\[0.3em]
    NGC 5904  & $7.9 \pm 0.5$  & $3.67$ \\[0.3em]
    NGC 6205  & $8.9 \pm 0.4$  & $3.35$ \\[0.3em]
    NGC 6388  & $16.3 \pm 5.2$ & $5.18$ \\[0.3em]
    NGC 6440  & $14.0 \pm 1.4$ & $5.10$ \\[0.3em]
    NGC 6441  & $17.6 \pm 2.0$ & $4.95$ \\[0.3em]
    NGC 6626  & $10.4 \pm 0.9$ & $4.98$ \\[0.3em]
    Terzan 5  & $15.6 \pm 1.2$ & $4.72$ \\[0.5em]
    \hline
    \multicolumn{3}{l}{\textit{Core-collapsed sample}} \\
    \hline
    NGC 362   & $8.6 \pm 0.4$  & $4.57$ \\[0.3em]
    NGC 6517  & $9.2 \pm 1.7$  & $5.63$ \\[0.3em]
    NGC 6624  & $7.0 \pm 0.4$  & $5.25$ \\[0.3em]
    NGC 6681  & $6.5 \pm 0.3$  & $6.03$ \\[0.3em]
    NGC 6752  & $7.7 \pm 0.4$  & $5.07$ \\[0.3em]
    NGC 7078  & $13.1 \pm 0.8$ & $6.65$ \\[0.3em]
    Terzan 1  & $9.8 \pm 0.9$  & $4.80$ \\
    \hline
    \multicolumn{3}{l}{\textit{Outliers}} \\
    \hline
    NGC 6266  & $14.8 \pm 1.6$ & $5.93$ \\[0.3em]
    NGC 6093* & $11.1 \pm 0.4$ & $5.52$ \\[0.3em]
    NGC 7089  & $11.1 \pm 0.6$ & $4.46$ \\
    \hline
  \end{tabular}
\end{table}

A caveat of our analysis is that several GCs have relatively few detected MSPs ($N \leq 10$, shown as blue points in Figure~\ref{fig:Frac_x_a_H}), limiting the statistical robustness of the inferred fractions. Moreover, we implicitly assume that the currently observed MSP samples are representative of the underlying populations, although observational selection effects, such as sensitivity limits, acceleration smearing, and eclipses, may bias the detected ratio of isolated to binary MSPs.

We excluded all core-collapsed GCs \citep[e.g., see][]{harris1996_GCs,beccari2006dynamical,yin2024-ngc6517_is_CC} from the fit shown in Figure \ref{fig:Frac_x_a_H} (see below for more details on the fit). With the exception of NGC 362, which follows the $1/a_H$ trend remarkably well and is analyzed separately below, these clusters occupy a distinct locus at the top of the $\mathcal{F}_i \times a_H$ plot (orange points in Figure~\ref{fig:Frac_x_a_H}), each harbouring more than five detected MSPs and showing an isolated-MSP fraction $\mathcal{F}_i > 80\%$. The classification of NGC 362 as core-collapsed remains debated\footnote{This cluster is classified only as a candidate core-collapsed in the 2010 version of the GLOBCLUST catalog: \url{https://heasarc.gsfc.nasa.gov/w3browse/all/globclust.html}}, and our data are more consistent with it being a non-core-collapsed GC.

\begin{figure}
    \centering
    \includegraphics[width=\linewidth]{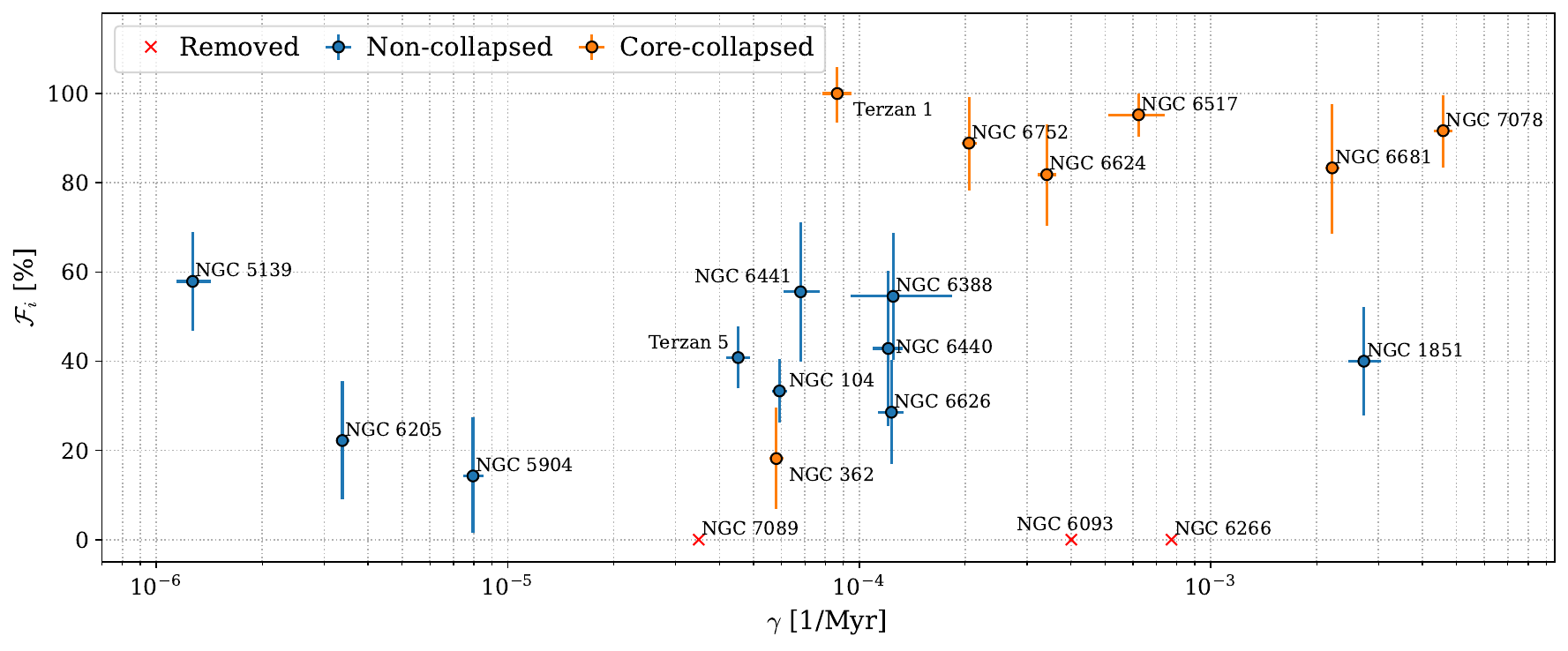}
    \includegraphics[width=\linewidth]{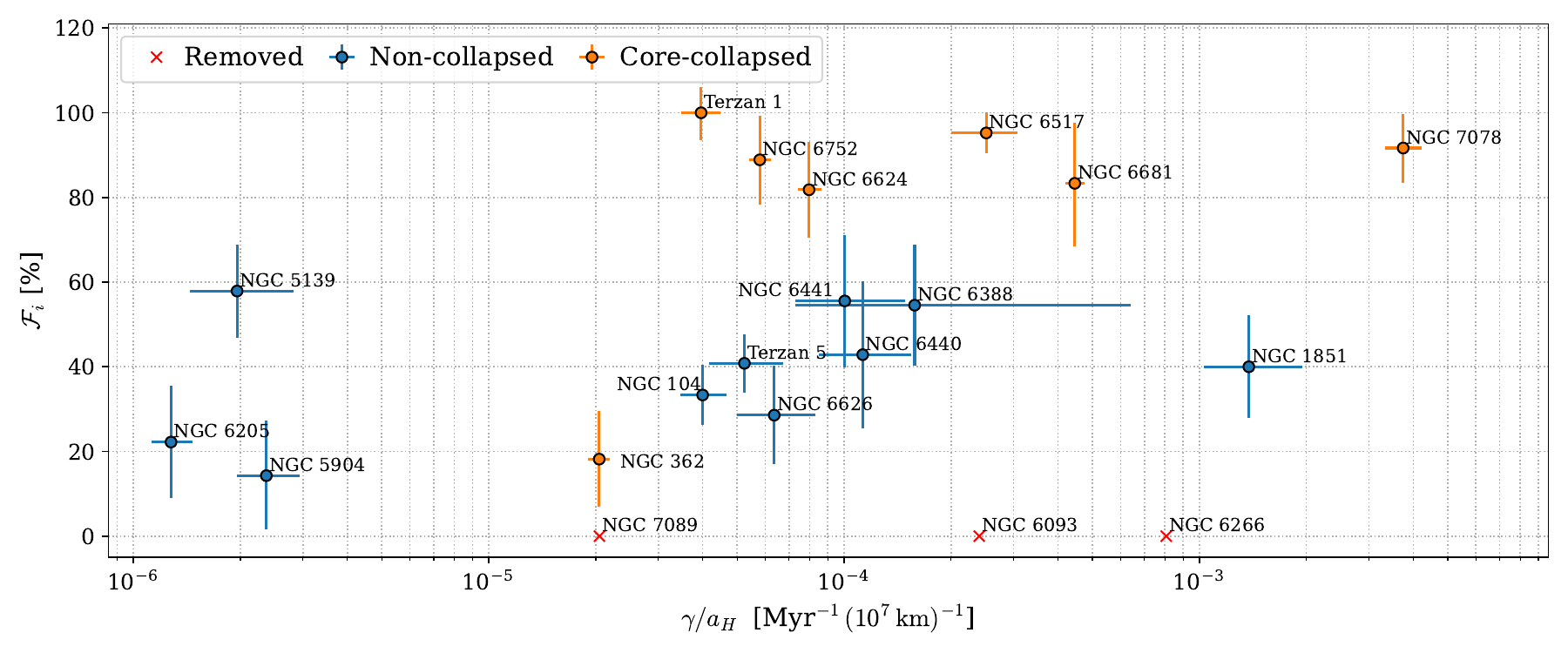}
    \caption{Fraction of isolated MSPs in GCs as a function of the encounter rate per single binary ($\gamma$, upper panel), and as a function of the ratio $\gamma/a_H$ (lower panel). No clear correlation is seen considering the full GC sample or the subsamples of core-collapsed and non-core-collapsed. These plots indicate that $\gamma$ is not a good predictor of the observed fraction of isolated MSPs in GCs. The most compelling case in both plots is that of NGC 5139 ($\omega$ Centauri), which has a relatively high $\mathcal{F}_i$ and very small $\gamma$.}
    \label{fig:enc_rate}
\end{figure}

To investigate the behavior of the core-collapsed clusters, in the upper panel of Figure~\ref{fig:enc_rate} we plot $\mathcal{F}_i$ against the encounter rate per binary $\gamma$ (Eq.~\ref{eq:enc_rate_per_binary}). We assume $M_{\rm tot} = 1.9\,M_\odot$ (approximately the sum of the MSP, companion, and incident-star masses) and a binary separation $a \approx 3\times 10^6$\,km, which corresponds to the median Keplerian separation of the 117 binary MSPs with measured orbital periods in the 20 GCs of our sample. The central number density is computed as $n = \rho_c/(0.4\,M_\odot)$, using the Baumgardt core-density estimates listed in Table~\ref{tab:velocity_dispersion}. The core-collapsed clusters tend to show relatively high $\gamma$ values, but no clear trend with $\mathcal{F}_i$ is apparent. We note that the role of a high $\gamma$ in binary ionization depends on which side of $a_H$ the binaries lie on, so a more reliable ionization indicator should combine both quantities, with ionization increasing if $\gamma$ is high and $a_H$ is small. In the lower panel of Figure~\ref{fig:enc_rate}, we therefore plot $\mathcal{F}_i$ against $\gamma/a_H$. Neither panel shows a clear trend, suggesting that other formation channels for isolated MSPs may be at work in core-collapsed clusters, as discussed in Section~\ref{sec:intro}.

We also identify three outliers with no isolated MSPs detected in our sample: NGC 6093, NGC 6266, and NGC 7089. These are marked with red crosses in Figures~\ref{fig:Frac_x_a_H} and \ref{fig:enc_rate}, where they are problematic in all three panels. They are excluded from the fitted models shown in Figure~\ref{fig:Frac_x_a_H}, but are retained for our three trend analyses, which are the Pearson linear correlation coefficient and the Spearman and Kendall monotonic correlation tests, shown in Table~\ref{tab:correlation_tests}, where we also report the correlation coefficients with NGC 362 included as a non-core-collapsed GC. All three tests indicate a very strong anti-correlation between $\mathcal{F}_i$ and $a_H$ in our main sample (MS) of 10 non-core-collapsed GCs, with correlation coefficients around $-0.9$ and p-values $2.3\times10^{-5}\leq p\leq 1.1\times10^{-4}$. Adding NGC 362 to the MS slightly strengthens these correlations, whereas adding the outliers substantially weakens their significance (although not refuting them entirely), yielding p-values in the range $8.2\times10^{-3}\leq p\leq 8.2\times10^{-2}$.

\begin{table}[h]
    \centering
    \begin{tabular}{lccc}
       \hline
       Sample & Pearson & Spearman & Kendall \\
       \hline
       MS  & $-0.93$ ($p=1.1\times10^{-4}$) & $-0.95$ ($p=2.3\times10^{-5}$) & $-0.87$ ($p=1.1\times10^{-4}$) \\
       MS + NGC 362  & $-0.94$ ($p=1.7\times10^{-5}$) & $-0.96$ ($p=2.1\times10^{-6}$) & $-0.89$ ($p=1.4\times10^{-5}$) \\
       MS + outliers & $-0.50$ ($p=8.2\times10^{-2}$) & $-0.65$ ($p=1.6\times10^{-2}$) & $-0.54$ ($p=1.2\times10^{-2}$)\\
       MS + outliers + NGC 362 & $-0.51$ ($p=6.1\times10^{-2}$) & $-0.63$ ($p=1.5\times10^{-2}$) & $-0.54$ ($p=8.2\times10^{-3}$)\\
    \end{tabular}
    \caption{Correlation coefficients for the main sample (MS) and for variations of this sample that include NGC 362 and/or the outliers. The MS and MS + NGC 362 samples show very strong monotonic anti-correlations, with p-values in the range $10^{-6}\lesssim p\lesssim 10^{-4}$. Including the outliers still yields evidence for an anti-correlation, though at substantially lower significance, with p-values in the range $8.2\times10^{-3}\leq p\leq 8.2\times10^{-2}$.}
    \label{tab:correlation_tests}
\end{table}

We employed the orthogonal distance regression (ODR) method from the \texttt{scipy} \citep{virtanen2020scipy} \texttt{Python} library to fit several power-law models to our main sample of 10 non-core-collapsed GCs, accounting for uncertainties in both axes. The results are shown in Figure~\ref{fig:Frac_x_a_H} and Table~\ref{tab:odr_fits}. We compared these models using a Bayes factor analysis, computing the marginal likelihood of each model via direct numerical integration over its free parameters ($A$ and, where applicable, $C$), adopting broad, proper uniform priors, $A \in (-300, 300)$ and $C \in (-100, 200)$, and confirming that our conclusions are stable under a widening or narrowing of these ranges. Because $\mathcal{F}_i$ is a binomial proportion, its uncertainty depends on the choice of prior for the binomial parameter; we therefore repeated the full Bayes factor analysis under both a Jeffreys prior and a uniform prior on $\mathcal{F}_i$. The resulting logarithms of the Bayes factors under each of these two priors are listed in the last two columns of Table~\ref{tab:odr_fits}, computed relative to the null hypothesis $\mathcal{F}_i = 38.90 \pm 3.56\%$, a constant fit with no dependence of $\mathcal{F}_i$ on $a_H$. The only model that shows strong evidence (i.e., $\ln(\mathrm{BF}) > 2.5$) against the null hypothesis under both priors is the power law $A\,a_H^{-0.5}$, which we therefore adopt as our fiducial model.

\begin{table}[h]
\centering
\begin{tabular}{lccccc}
\hline\hline
Model ($\mathcal{F}_i$) & $A$ & $C$ & $\chi^2_{\rm red}$ & $\ln\mathrm{BF}_{\rm Jeff}$ & $\ln\mathrm{BF}_{\rm Unif}$ \\
\hline
\text{*}$A\,a_H^{-0.5}$            & $42.49 \pm 1.62$  & --                 & 0.166 & 3.319  & 2.602  \\ 
\text{*}$A\,a_H^{-1}$              & $43.28 \pm 3.58$  & --                 & 0.522 & 0.759  & $-0.617$ \\
$A\,a_H^{-0.5}+C$          & $45.09 \pm 7.29$  & $-2.45 \pm 6.66$   & 0.183 & 1.232  & 0.559  \\
$A\,a_H^{-1}+C$            & $24.44 \pm 4.14$  & $17.19 \pm 3.71$   & 0.180 & 0.414  & $-0.198$ \\
\text{*}$A\,a_H+C$                 & $-12.33 \pm 2.60$ & $56.63 \pm 4.17$   & 0.305 & $-0.046$ & $-0.816$ \\
Null ($\mathcal{F}_i=C$)   & --                & $38.90 \pm 3.56$   & 1.100 & 0      & 0      \\
\hline\hline
\end{tabular}
\caption{ODR fitting results for the tested models, together with a constant null-hypothesis model, where $a_H$ is in units of $10^7$~km. The last two columns list the natural-log Bayes factor of each model relative to the null hypothesis, computed under a Jeffreys prior and a uniform prior on $\mathcal{F}_i$, respectively. The pure power law $A\,a_H^{-0.5}$ is the only model that shows strong evidence (i.e., $\ln(\mathrm{BF}) > 2.5$) against the null hypothesis under both priors. Models marked with an asterisk ($*$) are shown in Figure~\ref{fig:Frac_x_a_H}.}
\label{tab:odr_fits}
\end{table}

We further tested the stability of the fiducial-model fit with a leave-one-out (jackknife) analysis for four different data samples (see Table \ref{tab:jackknife}). The fit is very stable, with $A\approx42$, for both the main sample and the main sample plus NGC 362; in both cases, the cluster with the largest individual influence on the fit is Terzan 5. Once the three outliers are included, the jackknife fits become substantially less stable and the power law $A\,a_H^{-0.5}$ no longer describes the data well ($\chi^2_{\rm red} > 5$). The most influential cluster in both outlier-inclusive samples is NGC 6266: removing it from the main sample~+~outliers~+~NGC~362 changes the fit to $A = 26.79\pm5.27$. We also note that the Bayes factor against the null hypothesis is higher for the outlier-inclusive samples than for the main sample (Table~\ref{tab:jackknife}); this reflects the fact that the constant (null) model provides an even worse description of these three points with $\mathcal{F}_i=0\%$ than our fiducial power law. Although our fiducial model still describes the data better than the null hypothesis when the outliers are included, we exclude them from the main fit given the substantially poorer fit quality and the strong sensitivity of the result to a single cluster in their presence.

\begin{table}[h]
    \centering
    \begin{tabular}{lccccccc}
       \hline
       Sample & $A$ (full-sample fit) & $\chi^2_{\rm red}$ & $\langle A\rangle_{\rm JK}$ & $\ln\mathrm{BF}_{\rm Jeff}$ & $\ln\mathrm{BF}_{\rm Unif}$ & Most influential GC \\
       \hline
       MS                        & $42.49 \pm 1.62$ & 0.17 & $42.53 \pm 0.75$ & 3.319 & 2.602 & Terzan 5 \\
       MS + NGC 362               & $42.18 \pm 1.58$ & 0.16 & $42.21 \pm 0.65$ & 4.596 & 3.729 & Terzan 5 \\
       MS + outliers              & $19.65 \pm 6.24$ & 5.96 & $19.75 \pm 2.93$ & 6.737 & 6.571 & NGC 6266 \\
       MS + outliers + NGC 362    & $19.94 \pm 5.97$ & 5.55 & $20.04 \pm 2.77$ & 6.627 & 6.547 & NGC 6266 \\
       \hline
    \end{tabular}
    \caption{Jackknife analysis of the fiducial model, $A\,a_H^{-0.5}$, for our four data samples. The second column gives the best-fit amplitude $A$ using the full sample; the third gives its reduced $\chi^2$. The fourth column reports the mean and standard deviation of $A$ across the jackknife refits ($\langle A\rangle_{\rm JK}$). The fifth and sixth columns give the log Bayes factor of the fiducial model relative to the null hypothesis, computed under a Jeffreys prior and a uniform prior on $\mathcal{F}_i$, respectively. The seventh column identifies the cluster whose removal shifts $A$ the most.}
    \label{tab:jackknife}
\end{table}

\section{Discussion and conclusions}
\label{sec:conclusions}

In this work, we have proposed that the combination of the Heggie--Hills law for hard/soft binaries with companion ablation by the MSP gives us a robust indicator of the ionization fraction of binary MSPs in non-core-collapsed GCs. Our results provide a coherent explanation for a long-standing puzzle regarding the formation of isolated MSPs in GCs. Perhaps the most compelling case is that of NGC 5139 ($\omega$ Centauri), whose large MSP population (and particularly its high fraction of isolated MSPs) has consistently challenged theoretical explanations \citep[see Section 4.2 in][]{chen2023meerkat}. The framework presented here naturally accounts for the overabundance of isolated MSPs in this cluster.

Adopting the fiducial masses $M_{\rm NS} = 1.5$ M$_\odot$, $M_c = 0.01$ M$_\odot$, and $m_\ast = 0.4$ M$_\odot$, our best-fit model (Figure~\ref{fig:Frac_x_a_H} and Table \ref{tab:odr_fits}) predicts that all binary MSPs in a non-core-collapsed GC would be rapidly ionized when $a_H \approx 0.2 \times 10^7$ km, corresponding to $\mathcal{F}_i = 100\%$. This threshold translates to a central velocity dispersion of $\sigma \approx 33$ km s$^{-1}$, which exceeds the estimated values for any Milky Way GC. Adopting alternative fiducial masses (e.g., $M_c = 0.1$~M$_\odot$) does not, in practice, change the power-law scaling shown in Figure~\ref{fig:Frac_x_a_H}, although it does shift the corresponding values of $a_H$ systematically higher. This shift, however, would challenge the proposed explanation for the origin of isolated MSPs: the orbital separations of the currently observed compact binaries would then fall roughly 25--200 times below $a_H$, placing them closer to the Heggie--Hills hardening regime rather than the ionization regime. Even in this most pessimistic scenario, our results still show that non-core-collapsed clusters with smaller $a_H$ host higher fractions of isolated MSPs, so $a_H$ remains a useful empirical predictor of the binary MSP ionization fraction regardless of the precise fiducial masses adopted.

Interestingly, the fate of the liberated low-mass companions is likely dynamical evaporation from the cluster. Due to their extremely low masses and consequently high ejection speeds, these objects are unlikely to be recaptured into new binary systems via subsequent stellar encounters. Instead, they will diffuse toward the cluster outskirts and may ultimately be ejected into the Galactic halo. The isolated MSPs themselves can later share the same fate if they undergo gravitational interactions with more massive binaries, such as stellar-mass black hole binaries, or disrupt a main-sequence star in a close flyby \citep{kremer2022formation_of_low-mass}.

\begin{acknowledgments}
I dedicate this work to the memory of Prof. Flavio Menezes de Aguiar, my very first supervisor, who passed away in 2026. Flavio was a brilliant physicist and one of the kindest people I have ever worked with. Dear Professor, part of your knowledge was passed on to me, and now this legacy is being passed to my students. Thank you for showing me the path of science. I would also like to thank the anonymous referee for
the insightful comments and suggestions, which helped me to improve the quality of this manuscript substantially.
\end{acknowledgments}

%

\software{scipy \citep{virtanen2020scipy}
          }



\bibliography{sample701}{}

@PREAMBLE{
 "\providecommand{\noopsort}[1]{}" 
 # "\providecommand{\singleletter}[1]{#1}%" 
}

@article{chen2023meerkat,
  title={MeerKAT discovery of 13 new pulsars in Omega Centauri},
  author={Chen, Weiwei and Freire, PCC and Ridolfi, A and Barr, ED and Stappers, B and Kramer, M and Possenti, A and Ransom, SM and Levin, L and Breton, RP and others},
  journal={Monthly Notices of the Royal Astronomical Society},
  volume={520},
  number={3},
  pages={3847--3856},
  year={2023},
  publisher={Oxford University Press}
}

@article{deMenezes2023dynamical,
  title={How the dynamical properties of globular clusters impact their $\gamma$-ray and X-ray emission},
  author={de Menezes, Raniere and Di Pierro, Federico and Chiavassa, Andrea},
  journal={Monthly Notices of the Royal Astronomical Society},
  volume={523},
  number={3},
  pages={4455--4467},
  year={2023},
  publisher={Oxford University Press}
}

@article{deMenezes2019milky,
  title={Milky Way globular clusters in $\gamma$-rays: analysing the dynamical formation of millisecond pulsars},
  author={de Menezes, Raniere and Cafardo, Fabio and Nemmen, Rodrigo},
  journal={Monthly Notices of the Royal Astronomical Society},
  volume={486},
  number={1},
  pages={851--867},
  year={2019},
  publisher={Oxford University Press}
}

@article{verbunt1987formation,
  title={Formation of isolated millisecond pulsars in globular clusters},
  author={Verbunt, F and Van den Heuvel, EPJ and Van Paradijs, J and Rappaport, SA},
  journal={Nature},
  volume={329},
  number={6137},
  pages={312--314},
  year={1987},
  publisher={Nature Publishing Group UK London}
}

@article{bahramian2013stellar,
  title={Stellar encounter rate in galactic globular clusters},
  author={Bahramian, Arash and Heinke, Craig O and Sivakoff, Gregory R and Gladstone, Jeanette C},
  journal={The Astrophysical Journal},
  volume={766},
  number={2},
  pages={136},
  year={2013},
  publisher={The American Astronomical Society}
}

@book{lorimer2005handbook,
  title={Handbook of pulsar astronomy},
  author={Lorimer, Duncan Ross and Kramer, Michael},
  volume={4},
  year={2005},
  publisher={Cambridge university press}
}

@article{sollima2017global,
  title={The global mass functions of 35 Galactic globular clusters: I. Observational data and correlations with cluster parameters},
  author={Sollima, Antonio and Baumgardt, Holger},
  journal={Monthly Notices of the Royal Astronomical Society},
  volume={471},
  number={3},
  pages={3668--3679},
  year={2017},
  publisher={Oxford University Press}
}

@article{goldreich1969pulsar,
  title={Pulsar electrodynamics},
  author={Goldreich, Peter and Julian, William H},
  journal={Astrophysical Journal, vol. 157, p. 869},
  volume={157},
  pages={869},
  year={1969}
}

@article{paresce2000globular,
  title={On the Globular Cluster Initial Mass Function below 1 M},
  author={Paresce, Francesco and De Marchi, Guido},
  journal={The Astrophysical Journal},
  volume={534},
  number={2},
  pages={870--879},
  year={2000}
}

@article{archibald2009radio,
  title={A radio pulsar/X-ray binary link},
  author={Archibald, Anne M and Stairs, Ingrid H and Ransom, Scott M and Kaspi, Victoria M and Kondratiev, Vladislav I and Lorimer, Duncan R and McLaughlin, Maura A and Boyles, Jason and Hessels, Jason WT and Lynch, Ryan and others},
  journal={Science},
  volume={324},
  number={5933},
  pages={1411--1414},
  year={2009},
  publisher={American Association for the Advancement of Science}
}

@book{frank2002accretion,
  title={Accretion power in astrophysics},
  author={Frank, Juhan and King, Andrew R and Raine, Derek},
  year={2002},
  publisher={Cambridge university press}
}

@article{pooley2003dynamical,
  title={Dynamical formation of close binary systems in globular clusters},
  author={Pooley, David and Lewin, Walter HG and Anderson, Scott F and Baumgardt, Holger and Filippenko, Alexei V and Gaensler, Bryan M and Homer, Lee and Hut, Piet and Kaspi, Victoria M and Makino, Junichiro and others},
  journal={The Astrophysical Journal Letters},
  volume={591},
  number={2},
  pages={L131--L134},
  year={2003}
}

@article{verbunt2014disruption,
  title={On the disruption of pulsar and X-ray binaries in globular clusters},
  author={Verbunt, Frank and Freire, Paulo CC},
  journal={Astronomy \& Astrophysics},
  volume={561},
  pages={A11},
  year={2014},
  publisher={EDP Sciences}
}

@article{katz1975two,
  title={Two kinds of stellar collapse},
  author={Katz, JI},
  journal={Nature},
  volume={253},
  number={5494},
  pages={698--699},
  year={1975},
  publisher={Nature Publishing Group UK London}
}

@article{smith2023third,
  title={The third Fermi Large Area Telescope catalog of gamma-ray pulsars},
  author={Smith, David A and Abdollahi, S and Ajello, M and Bailes, Matthew and Baldini, L and Ballet, J and Baring, MG and Bassa, C and Gonzalez, J Becerra and Bellazzini, R and others},
  journal={The Astrophysical Journal},
  volume={958},
  number={2},
  pages={191},
  year={2023},
  publisher={The American Astronomical Society}
}

@article{ye2019millisecond,
  title={Millisecond pulsars and black holes in globular clusters},
  author={Ye, Claire S and Kremer, Kyle and Chatterjee, Sourav and Rodriguez, Carl L and Rasio, Frederic A},
  journal={The Astrophysical Journal},
  volume={877},
  number={2},
  pages={122},
  year={2019},
  publisher={The American Astronomical Society}
}

@incollection{Kremer2026_compact_objects_in_GCs,
title = {Compact objects in globular clusters},
editor = {Ilya Mandel},
booktitle = {Encyclopedia of Astrophysics (First Edition)},
publisher = {Elsevier},
edition = {First Edition},
address = {Oxford},
pages = {458-472},
year = {2026},
isbn = {978-0-443-21440-0},
doi = {https://doi.org/10.1016/B978-0-443-21439-4.00103-6},
url = {https://www.sciencedirect.com/science/article/pii/B9780443214394001036},
author = {Kyle Kremer}
}

@article{dai2020discovery,
  title={Discovery of millisecond pulsars in the globular cluster omega centauri},
  author={Dai, Shi and Johnston, Simon and Kerr, Matthew and Camilo, Fernando and Cameron, Andrew and Toomey, Lawrence and Kumamoto, Hiroki},
  journal={The Astrophysical Journal Letters},
  volume={888},
  number={2},
  pages={L18},
  year={2020},
  publisher={The American Astronomical Society}
}

@article{evans2023gamma,
  title={On the gamma-ray emission from the core of the Sagittarius dwarf galaxy},
  author={Evans, Addy J and Strigari, Louis E and Svenborn, Oskar and Albert, Andrea and Harding, J Patrick and Hooper, Dan and Linden, Tim and Pace, Andrew B},
  journal={Monthly Notices of the Royal Astronomical Society},
  volume={524},
  number={3},
  pages={4574--4585},
  year={2023},
  publisher={Oxford University Press}
}

@article{heggie1975binary,
  title={Binary evolution in stellar dynamics},
  author={Heggie, Douglas C},
  journal={Monthly Notices of the Royal Astronomical Society},
  volume={173},
  number={3},
  pages={729--787},
  year={1975},
  publisher={Oxford University Press Oxford, UK}
}

@article{hills1975encounters,
  title={Encounters between binary and single stars and their effect on the dynamical evolution of stellar systems},
  author={Hills, JG},
  journal={Astronomical Journal, vol. 80, Oct. 1975, p. 809-825.},
  volume={80},
  pages={809--825},
  year={1975}
}

@article{gentile2014xray_black_widow,
  title={X-ray observations of black widow pulsars},
  author={Gentile, PA and Roberts, MSE and McLaughlin, MA and Camilo, F and Hessels, JWT and Kerr, M and Ransom, SM and Ray, PS and Stairs, IH},
  journal={The Astrophysical Journal},
  volume={783},
  number={2},
  pages={69},
  year={2014},
  publisher={The American Astronomical Society}
}

@article{hut1983binary,
  title={Binary-single star scattering. I-Numerical experiments for equal masses},
  author={Hut, Piet and Bahcall, John N},
  journal={Astrophysical Journal, Part 1 (ISSN 0004-637X), vol. 268, May 1, 1983, p. 319-341.},
  volume={268},
  pages={319--341},
  year={1983}
}

@article{romani2012psr,
  title={PSR J1311- 3430: A Heavyweight Neutron Star with a Flyweight Helium Companion},
  author={Romani, Roger W and Filippenko, Alexei V and Silverman, Jeffrey M and Cenko, S Bradley and Greiner, Jochen and Rau, Arne and Elliott, Jonathan and Pletsch, Holger J},
  journal={The Astrophysical Journal Letters},
  volume={760},
  number={2},
  pages={L36},
  year={2012},
  publisher={The American Astronomical Society}
}

@article{fruchter1988millisecond,
  title={A millisecond pulsar in an eclipsing binary},
  author={Fruchter, AS and Stinebring, DR and Taylor, JH},
  journal={Nature},
  volume={333},
  number={6170},
  pages={237--239},
  year={1988},
  publisher={Nature Publishing Group UK London}
}

@article{spiewak2018psr,
  title={PSR J2322- 2650--a low-luminosity millisecond pulsar with a planetary-mass companion},
  author={Spiewak, R and Bailes, Matthew and Barr, ED and Bhat, NDR and Burgay, MARTA and Cameron, AD and Champion, DJ and Flynn, CML and Jameson, A and Johnston, S and others},
  journal={Monthly Notices of the Royal Astronomical Society},
  volume={475},
  number={1},
  pages={469--477},
  year={2018},
  publisher={Oxford University Press}
}

@article{ginzburg2021black,
  title={Black widow formation by pulsar irradiation and sustained magnetic braking},
  author={Ginzburg, Sivan and Quataert, Eliot},
  journal={Monthly Notices of the Royal Astronomical Society},
  volume={500},
  number={2},
  pages={1592--1603},
  year={2021},
  publisher={Oxford University Press}
}

@article{devito2020identifying,
  title={Identifying the formation mechanism of redback pulsars},
  author={De Vito, Mar{\'\i}a Alejandra and Benvenuto, Omar Gustavo and Horvath, Jorge Ernesto},
  journal={Monthly Notices of the Royal Astronomical Society},
  volume={493},
  number={2},
  pages={2171--2177},
  year={2020},
  publisher={Oxford University Press}
}

@article{sollima2019eye,
  title={The eye of Gaia on globular clusters kinematics: internal rotation},
  author={Sollima, A and Baumgardt, H and Hilker, M},
  journal={Monthly Notices of the Royal Astronomical Society},
  volume={485},
  number={1},
  pages={1460--1476},
  year={2019},
  publisher={Oxford University Press}
}

@article{baumgardt2023evidence,
  title={Evidence for a bottom-light initial mass function in massive star clusters},
  author={Baumgardt, Holger and Henault-Brunet, Vincent and Dickson, Nolan and Sollima, Antonio},
  journal={Monthly Notices of the Royal Astronomical Society},
  volume={521},
  number={3},
  pages={3991--4008},
  year={2023},
  publisher={Oxford University Press}
}

@article{baumgardt2018catalogue,
  title={A catalogue of masses, structural parameters, and velocity dispersion profiles of 112 Milky Way globular clusters},
  author={Baumgardt, Holger and Hilker, Michael},
  journal={Monthly Notices of the Royal Astronomical Society},
  volume={478},
  number={2},
  pages={1520--1557},
  year={2018},
  publisher={Oxford University Press}
}

@article{ridolfi2022trapum,
  title={TRAPUM discovery of 13 new pulsars in NGC 1851 using MeerKAT},
  author={Ridolfi, Alessandro and Freire, PCC and Gautam, T and Ransom, SM and Barr, ED and Buchner, S and Burgay, M and Abbate, F and Krishnan, V Venkatraman and Vleeschower, L and others},
  journal={Astronomy \& Astrophysics},
  volume={664},
  pages={A27},
  year={2022},
  publisher={EDP Sciences}
}

@article{freire2017long,
  title={Long-term observations of the pulsars in 47 Tucanae--II. Proper motions, accelerations and jerks},
  author={Freire, PCC and Ridolfi, ALESSANDRO and Kramer, M and Jordan, C and Manchester, RN and Torne, P and Sarkissian, J and Heinke, CO and D’Amico, N and Camilo, F and others},
  journal={Monthly Notices of the Royal Astronomical Society},
  volume={471},
  number={1},
  pages={857--876},
  year={2017},
  publisher={Oxford University Press}
}

@article{padmanabh2024discovery,
  title={Discovery and timing of ten new millisecond pulsars in the globular cluster Terzan 5},
  author={Padmanabh, PV and Ransom, SM and Freire, PCC and Ridolfi, A and Taylor, JD and Choza, C and Clark, CJ and Abbate, F and Bailes, M and Barr, ED and others},
  journal={Astronomy \& Astrophysics},
  volume={686},
  pages={A166},
  year={2024},
  publisher={EDP Sciences}
}

@ARTICLE{harris1996_GCs,
       author = {{Harris}, William E.},
        title = "{A Catalog of Parameters for Globular Clusters in the Milky Way}",
      journal = {Astronomical Journal},
         year = 1996,
       volume = {112},
        pages = {1487},
          doi = {10.1086/118116},
       adsurl = {https://ui.adsabs.harvard.edu/abs/1996AJ....112.1487H}
}

@article{pan2021fast,
  title={FAST Globular Cluster Pulsar survey: twenty-four pulsars discovered in 15 globular clusters},
  author={Pan, Zhichen and Qian, Lei and Ma, Xiaoyun and Liu, Kuo and Wang, Lin and Luo, Jintao and Yan, Zhen and Ransom, Scott and Lorimer, Duncan and Li, Di and others},
  journal={The Astrophysical Journal Letters},
  volume={915},
  number={2},
  pages={L28},
  year={2021},
  publisher={The American Astronomical Society}
}

@ARTICLE{ye2024Single_Millisecond,
       author = {{Ye}, Claire S. and {Kremer}, Kyle and {Ransom}, Scott M. and {Rasio}, Frederic A.},
        title = "{Single Millisecond Pulsars from Dynamical Interaction Processes in Dense Star Clusters}",
      journal = {The Astrophysical Journal},
         year = 2024,
        month = jan,
       volume = {961},
       number = {1},
          eid = {98},
        pages = {98},
          doi = {10.3847/1538-4357/ad089a},
archivePrefix = {arXiv},
       eprint = {2307.15740},
 primaryClass = {astro-ph.HE},
       adsurl = {https://ui.adsabs.harvard.edu/abs/2024ApJ...961...98Y}
}

@ARTICLE{kremer2022formation_of_low-mass,
       author = {{Kremer}, Kyle and {Ye}, Claire S. and {K{\i}ro{\u{g}}lu}, Fulya and {Lombardi}, James C. and {Ransom}, Scott M. and {Rasio}, Frederic A.},
        title = "{Formation of Low-mass Black Holes and Single Millisecond Pulsars in Globular Clusters}",
      journal = {Astrophys. J. Lett},
         year = 2022,
        month = jul,
       volume = {934},
       number = {1},
          eid = {L1},
        pages = {L1},
          doi = {10.3847/2041-8213/ac7ec4},
archivePrefix = {arXiv},
       eprint = {2204.07169},
 primaryClass = {astro-ph.HE},
       adsurl = {https://ui.adsabs.harvard.edu/abs/2022ApJ...934L...1K}
}

@ARTICLE{ye2024white_dwarf_collision,
       author = {{Ye}, Claire S. and {Kremer}, Kyle and {Ransom}, Scott M. and {Rasio}, Frederic A.},
        title = "{Single Millisecond Pulsars from Dynamical Interaction Processes in Dense Star Clusters}",
      journal = {Astrophys. J.},
         year = 2024,
        month = jan,
       volume = {961},
       number = {1},
          eid = {98},
        pages = {98},
          doi = {10.3847/1538-4357/ad089a},
archivePrefix = {arXiv},
       eprint = {2307.15740},
 primaryClass = {astro-ph.HE},
       adsurl = {https://ui.adsabs.harvard.edu/abs/2024ApJ...961...98Y}
}

@INPROCEEDINGS{camilo2005psr_in_gc,
       author = {{Camilo}, F. and {Rasio}, F.~A.},
        title = "{Pulsars in Globular Clusters}",
    booktitle = {Binary Radio Pulsars},
         year = 2005,
       editor = {{Rasio}, Fred A. and {Stairs}, Ingrid H.},
       series = {Astronomical Society of the Pacific Conference Series},
       volume = {328},
        month = jul,
        pages = {147},
          doi = {10.48550/arXiv.astro-ph/0501226},
archivePrefix = {arXiv},
       eprint = {astro-ph/0501226},
 primaryClass = {astro-ph},
       adsurl = {https://ui.adsabs.harvard.edu/abs/2005ASPC..328..147C}
}

@ARTICLE{ye2022_compact,
       author = {{Ye}, Claire S. and {Kremer}, Kyle and {Rodriguez}, Carl L. and {Rui}, Nicholas Z. and {Weatherford}, Newlin C. and {Chatterjee}, Sourav and {Fragione}, Giacomo and {Rasio}, Frederic A.},
        title = "{Compact Object Modeling in the Globular Cluster 47 Tucanae}",
      journal = {Astrophys. J.},
         year = 2022,
        month = jun,
       volume = {931},
       number = {2},
          eid = {84},
        pages = {84},
          doi = {10.3847/1538-4357/ac5b0b},
archivePrefix = {arXiv},
       eprint = {2110.05495},
 primaryClass = {astro-ph.HE},
       adsurl = {https://ui.adsabs.harvard.edu/abs/2022ApJ...931...84Y}
}

@ARTICLE{Sigurdsson1995Dynamics,
       author = {{Sigurdsson}, Steinn and {Phinney}, E.~S.},
        title = "{Dynamics and Interactions of Binaries and Neutron Stars in Globular Clusters}",
      journal = {Astrophys. J. Sup. Ser.},
         year = 1995,
        month = aug,
       volume = {99},
        pages = {609},
          doi = {10.1086/192199},
archivePrefix = {arXiv},
       eprint = {astro-ph/9412078},
 primaryClass = {astro-ph},
       adsurl = {https://ui.adsabs.harvard.edu/abs/1995ApJS...99..609S}
}

@ARTICLE{ivanova2008Formation,
       author = {{Ivanova}, N. and {Heinke}, C.~O. and {Rasio}, F.~A. and {Belczynski}, K. and {Fregeau}, J.~M.},
        title = "{Formation and evolution of compact binaries in globular clusters - II. Binaries with neutron stars}",
      journal = {Month. Not. of the Royal Ast. Soc.},
         year = 2008,
        month = may,
       volume = {386},
       number = {1},
        pages = {553-576},
          doi = {10.1111/j.1365-2966.2008.13064.x},
archivePrefix = {arXiv},
       eprint = {0706.4096},
 primaryClass = {astro-ph},
       adsurl = {https://ui.adsabs.harvard.edu/abs/2008MNRAS.386..553I}
}

@article{virtanen2020scipy,
  title={SciPy 1.0: fundamental algorithms for scientific computing in Python},
  author={Virtanen, Pauli and Gommers, Ralf and Oliphant, Travis E and Haberland, Matt and Reddy, Tyler and Cournapeau, David and Burovski, Evgeni and Peterson, Pearu and Weckesser, Warren and Bright, Jonathan and others},
  journal={Nature methods},
  volume={17},
  number={3},
  pages={261--272},
  year={2020},
  publisher={Nature Publishing Group US New York}
}

@article{yin2025illuminating,
  title={Illuminating Hidden Pulsars: Scintillation-enhanced Discovery of Two Binary Millisecond Pulsars in M13 with FAST},
  author={Yin, Dejiang and Wang, Lin and Zhang, Li-yun and Qian, Lei and Li, Baoda and Liu, Kuo and Peng, Bo and Dai, Yinfeng and Li, Yaowei and Pan, Zhichen},
  journal={The Astrophysical Journal},
  volume={991},
  number={2},
  pages={177},
  year={2025},
  publisher={The American Astronomical Society}
}

@article{yin2024-ngc6517_is_CC,
  title={FAST discovery of eight isolated millisecond pulsars in NGC 6517},
  author={Yin, Dejiang and Zhang, Li-yun and Qian, Lei and Eatough, Ralph P and Li, Baoda and Lorimer, Duncan R and Dai, Yinfeng and Li, Yaowei and Zhang, Xingnan and Li, Minghui and others},
  journal={The Astrophysical Journal Letters},
  volume={969},
  number={1},
  pages={L7},
  year={2024},
  publisher={The American Astronomical Society}
}

@article{beccari2006dynamical,
  title={The dynamical state and blue straggler population of the globular cluster NGC 6266 (M62)},
  author={Beccari, G and Ferraro, Francesco Rosario and Possenti, A and Valenti, Elena and Origlia, L and Rood, RT},
  journal={The Astronomical Journal},
  volume={131},
  number={5},
  pages={2551--2560},
  year={2006}
}

@article{oh2023influences,
  title={Influences of dynamical disruptions on the evolution of pulsars in globular clusters},
  author={Oh, Kwangmin and Hui, CY and Hong, Jongsuk and Takata, J and Kong, AKH and Tam, Pak-Hin Thomas and Li, Kwan-Lok and Cheng, KS},
  journal={Monthly Notices of the Royal Astronomical Society},
  volume={525},
  number={3},
  pages={4167--4175},
  year={2023},
  publisher={Oxford University Press}
}
\bibliographystyle{aasjournalv7}



\end{document}